\documentclass[10pt,english,aps,prl,showpacs,twocolumn]{revtex4}
\makeatletter
\usepackage{graphicx}
\usepackage{epsfig}
\usepackage{babel}

\newcommand{\beq}{\begin{eqnarray}}
\newcommand{\eeq}{\end{eqnarray}}
\newcommand{\n}{\nonumber}

\begin{document}

\title
{Universal sound absorption in low-temperature glasses}

\author{Misha  Turlakov}
\affiliation{Cavendish Laboratory, University of Cambridge, Cambridge, CB3 OHE, UK}
\date{\today}

\begin{abstract}
  
 Phenomenologically assuming a sharp decrease of shear relaxation
time for large wavevector $k>k_\xi$  density modes (where $k_\xi$ is of order of inverse of several interatomic distances $a$),
I develop a general elasto-hydrodynamic theory describing the low-energy excitations 
of glassy and amorphous solids, which are responsible for anomalous specific heat and thermal conductivity.
The theory explains the origin of collective two-level states and Boson peak.
The ratio of the wavelength of the phonon, $\lambda$, to its mean free path, $l$, - universal property
of sound absorption in glasses - is derived in this theory to be $\lambda/l=(2/3) (c_t/c_l)^2 (k_\xi a)^3$,
where $c_t$ and $c_l$ are transverse and longitudinal sound velocities correspondingly.

\end{abstract}
\maketitle


Glasses at low temperatures ($T < 1-5~K$) show a remarkable almost universal sound absorption -
``the ratio of the phonon wavelength, $\lambda$, to the phonon mean free path, $l$, has been
found to lie between $10^{-3}$ and $10^{-2}$ in almost all cases, independent
of chemical composition and frequency (wavelength) of the elastic wave which varied by more
than nine orders of magnitude in the different experiments''\cite{Pohl}. 
The so called Standard Tunneling Model of glasses\cite{Anderson} assumes phenomenologically
the existence of {\it independent local tunneling two-level systems} (tunneling TLS's) with
wide distribution of tunneling splittings and relaxation times. The Standard Model had a lot of successes
in describing many low-temperature properties of glasses (specific heat, thermal conductivity, etc.)
(for a review, see \cite{Phillips}).
The {\it coherent and weakly interacting nature} of excitations was revealed by echo and
spectral diffusion experiments,
while the {\it tunneling aspect} of TLS's (the low-energy states which are due to quantum tunneling splitting
of local potential minima states of atoms (or group of atoms)) was never clearly demonstrated experimentally\cite{defense}.
Most importantly, the understanding of universal sound absorption is beyond the scope
of the Standard Model, because the density of tunneling TLS's has to be assumed {\it arbitrarily}
for this model\cite{Leggett}. Moreover, the {\it independence of tunneling TLS's} implies large variation
and non-uniqueness of density of TLS's in different materials in contradiction to experimental
observations\cite{Pohl}.

In this Letter I construct a phenomenological  visco-elastic (or hydrodynamic) theory  with a single
assumption that a structural shear modulus relaxation time has a sharp threshold and becomes
finite for large-wavevector elastic perturbations(waves) of amorphous structure (this assumption is
reminiscent of Maxwell theory of viscoelastic liquids with a difference that here the relaxation time
is assumed to become finite only for short-wavelength perturbations). The low-energy states of this theory
(in addition to phonons), contributing an almost linear-temperature specific heat, are collective
density modes with large wavevectors $k$ and small frequencies. The same assumption leads
naturally to the appearance (or rather redistribution) of additional density of states at high frequencies
(so called Boson peak). The universal dimensionless constant $\lambda/l$ of amorphous condensed matter
physics is understood in this theory due to the weak coupling between phonons (small $q$) and 
relaxational collective (large $q$) modes.

There is more to be said about universality of $\lambda/l$ ratio in glassy and other types of materials.
It appears that at low temperatures only for good crystalline solids and quantum liquids ($He^3$ and $He^4$) this
universality is not applicable (for few other exceptions see a review\cite{Pohl}), while for
a huge class of materials\cite{Pohl} - large number of disordered crystals and polycrystals,
some quasicrystals and metallic glasses in addition to insulating glasses of various types -
the ratio of $\lambda/l$ falls into the same range $10^{-3}$ to $10^{-2}$. Another insightful
experimental observation is that the irradiation
of crystalline silicon saturates such a material ultimately with increasing dose of irradiation
 to the same universality\cite{Liu}.
It is illogical to believe that {\it independent tunneling} states are generated at a fixed universal
density. It seems more general assumption is required to understand this remarkable universality, since
neither amorphicity (lack of long-range order) nor glassiness is a necessary condition
for universality to appear.

Glassy and amorphous solids are solids in the sense that shear and compressional moduluses are finite
at low frequencies.
The transverse and longitudinal phonons with long
wavelengths are well-defined excitations,
since the  phonon mean free path is much longer than its wavelength, $l \sim 10^2 \lambda$.
As the wavelength of phonon becomes comparable to the correlation length of frozen disorder\cite{remark1}, 
the anharmonicity cannot be neglected and amorphous structure can relax under shear strain.
Such a qualitative argument indicates a physical reason why a shear relaxation time can become
finite for short wavelength perturbations (rigorous justification of the assumption is the subject
of microscopic theory and beyond the scope of this Letter). To be precise, the main {\bf phenomenological  assumption}
(to be referred to PA) is the following statement:
{\it structural shear relaxation time decreases sharply as a function of the wavevector $k$
from infinity at small wavevectors ($k < k_\xi \equiv 1/\xi$, where $\xi$ is the typical correlation length
of disorder in glasses of order of several interatomic or intermolecular (or any elementary unit of structure) distances)
 to a finite relaxational time smaller than 
inverse phonon frequency for wavevectors $k > k_\xi$}. 
Naturally, excitations for $k>1/\xi$ cannot be described as well defined excitations. Another way to
substantiate the assumption is to compare various phonon scattering mechanisms. Elastic Rayleigh scattering
of phonons by disorder has strong dependence on the wavelength $\lambda$ of phonon, at most
$\lambda/l=4\pi^2 (\xi/\lambda)^3$, where $\xi^3$ is the ``correlation volume'' of disorder.
Anharmonic three-phonon processes, being proportional to the phonon density of states (of high frequency
or thermal phonon involved), become also
more relevant at higher frequencies (or shorter wavelengths). 
A combined effect of disorder
and anharmonicity can conspire to the sharp threshold and drop to the finite shear relaxation time,
describing the slow rearrangement of atoms.
The main idea of the paper is to construct a minimal phenomenological hydrodynamic theory based
only on macroscopic conservation laws and the phenomenological assumption.

To derive the spectrum of density modes it is sufficient to use macroscopic conservation laws
and constitutive relation for the stress tensor\cite{Forster,Landau}. 
The conservation of the mass density $\rho$ and the conservation of momentum density momentum $\rho v_i$ are
in a linearized approximation
\beq
\partial_t \rho + \rho_0 div \vec{v}=0, \label{eqn:density} \\
\rho_0 \partial_t v_i -\nabla_j \sigma_{ij}=-n_0 \nabla_i \delta V_{ext}, 
\label{eqn:momentum}
\eeq
where $v_i$ and $\sigma_{ij}$ are local values of velocity and stress tensor fields, $n_0=\rho_0/m$ is
an equilibrium number density. The constitutive relation between the stress tensor $\sigma_{ij}$
and the strain tensor $u_{ij}$ is assumed to be
\beq
\sigma_{ij}=\frac{2\mu}{1+i/(\omega \tau_s (k))} u_{ij} +\Lambda u_{ll} \delta_{ij},
\label{eqn:constitute}
\eeq
where $\Lambda$ and $\mu$ are Lame coefficients\cite{Landau}, and $\tau_s (k)$ is the shear
relaxation time, dependent on the wavevector $k$ of the periodic modulation. Notice that
if $\omega \tau_s (k) \gg 1$, the constitutive relation describes a solid body with a finite
shear modulus. While for $\omega \tau_s (k) \ll 1$, the constitutive relation describes
a liquid with frequency- and wavevector-dependent viscosity $\eta_k (\omega)=\mu \tau_s(k)/(1-i\omega\tau_s(k))$.
Coupling between thermal variables (e.g. gradients of temperature) and mechanical variables
is neglected here. 
A particular functional dependence of $\tau_s(k)$ is not considered here.
This dependence may vary among various materials and can be used as a fitting parameter.
The only requirement is that $\tau_s(k)$ drops sufficiently fast for $k \sim k_\xi$,
and for larger $k \gg k_\xi$ the modes are relaxational so that $c_l k \tau_s(k) \ll 1$, where
$c_l$ is the longitudinal sound velocity. Substituting Eqns.(\ref{eqn:density}) and (\ref{eqn:constitute})
into Eqn.(\ref{eqn:momentum}), the longitudinal density correlation function can be calculated:
\beq
\chi_{\rho}(k,\omega)=\frac{n_0 k^2/m}{\omega^2-k^2(c^2_\infty-(c^2_\infty-c^2_0)/(1-i\omega\tau_s(k)))},
\label{eqn:spectrum}
\eeq
with $c_\infty^2 \equiv c_l^2=(\Lambda+2\mu)/\rho_0$ and $c_0^2=\Lambda/\rho_0$.
The longitudinal density fluctuation spectrum is given by $Im \chi_\rho (q,\omega)$
and has long low-frequency ($\omega \ll c_l k$) tails, which can be approximately written as
(assuming $c_l k \tau_s(k) \ll 1$):
\beq
Im \chi_\rho (k,\omega) \cong \frac{n_0}{m} \frac{c_\infty^2-c_0^2}{c_0^4} 
\frac{\omega \tau_s (k)}{1+(\omega \tau_s (k))^2 (c_\infty/c_0)^4}.
\label{eqn:low-freq}
\eeq
Therefore, due to shear stress relaxation, collective density modes at large $k > k_\xi$
have a {\it linear ``bosonic'' density of states at low frequencies}.
These modes of low frequency and short wavelength describe correlated in space local
rearrangements of atoms due to shear relaxation, and thus some aspects
of the microscopic nature of these modes are not very different from the Standard Model.
Such atomic rearrangements are strongly inhibited in a crystalline solid due to
the presence of short and long range order.

These modes contribute almost linear specific heat $C_V(T)$. The entropy
of interacting boson modes can be written as following\cite{Fulde}:
\beq
S(T) & \simeq & \frac{1}{\pi} {\displaystyle \int_0^\infty} d\omega \frac{e^{\beta \omega}}{(e^{\beta\omega}-1)^2}
\frac{\omega}{T^2} {\displaystyle \sum_k} \left[ Im ln \left( \frac{\chi_\rho (k,\omega)}{n_0/(mc_l^2)} \right) +  \right. \n \\
& + & \left. \frac{(c_\infty^2-c_0^2) k^2 \omega \tau_s (k)}{1+(\omega \tau_s (k))^2}
Re\left( \frac{\chi_\rho (k,\omega)}{n_0/(mc_l^2)} \right)  \right].  
\label{eqn:entropy} 
\eeq
For the purpose of approximate estimates, the specific heat $C_V (T)$ after integrating over the range
of wavevectors $k_\xi <k < k_u$ ($k_u$ is upper cutoff) with some typical relaxation time $\tau_{s,typ}$ is
\beq
C_V (T) \sim A (k_B T \tau_{s,typ}) \frac{k_u^3}{3\pi^2} \frac{c_\infty^2-c_0^2}{c_0^2},
\label{eqn:cv}
\eeq
where $A$ is a numerical factor. A crude estimate 
($\hbar/\tau_{s,typ} \sim \hbar \Omega_{BP} \equiv k_B T_{BP}$, $k_B T_{BP} =50~K $, $k_u \sim 10^7cm^{-1}$,
and a last factor as $0.3$) gives the density of states $\sim 10^{33} (erg cm^3)^{-1}$, which is consistent with experiment.
We eschew numerical factors and details, dependent
on certain latitude of the dependence of $\tau_s (k)$ in attempt to focus
on the essential consequences. The essential result for the linear specific heat
can be already seen from the linear frequency density of states for bosonic modes(Eqn.\ref{eqn:low-freq}),
 since the sum
over $k$ at low frequencies for the second term in Eqn.(\ref{eqn:entropy}) 
is approximately equal to $\sum_k Im \chi_\rho (k,\omega)$
from Eqn.(\ref{eqn:low-freq}).

Phenomenological assumption implies straightforwardly accumulation of additional density of states around frequency
$\Omega_{BP} \simeq c_0 k_\xi$, 
since $\tau_s(k)$ drops {\it sharply} to a finite value
around $k_\xi$. This is so called Boson peak observed in neutron and Raman scattering\cite{Benassi}. 
First, notice that the maximum of the spectrum $Im \chi_\rho (k,\omega)$ as a function of
frequency $\omega$ is shifted downward from $c_\infty k $ if $\omega \tau_s \gg 1$ 
toward $c_0 k$ if $\omega \tau_s \ll 1$. Since it happens
for the range of wavevectors $k$, large additional weight from the range of $k$-wavevectors will be redistributed from high frequencies
to the frequency around Boson peak. Second, the shape of the intensity $Im \chi_\rho (k,\omega)$
is asymmetric with larger weight above the maximum. All these factors contribute to the additional
density of states around $\Omega_{BP}$: $\nu (\omega) \sim \sum_k Im \chi_\rho (k,\omega \sim c_0 k_\xi)$.
The intensity of Boson peak would vary depending on a detailed dependence of $\tau_s (k)$, and, indeed,
experimentally intensity of Boson peak varies significantly among various glasses.

The calculation of sound attenuation\cite{thermal} needs to take into account a wide distribution of relaxation times
postulated by PA. We are interested in the attenuation of low-frequency phonon with frequency $\omega$ by 
slow relaxational density modes such that $\omega \tau_s(k) \ll 1$, and therefore hydrodynamic description of attenuation
is applicable. The simplest way to calculate
the sound absorption is to use a classical expression for absorption due to finite viscosity\cite{Landau}:
\beq
\gamma_l=\frac{\omega^2}{2\rho c_l^3} (4/3) Re \eta_k (\omega).
\eeq
The low-frequency phonon will be absorbed by all collective modes with various large wavevectors $k$
and corresponding relaxation times $\tau_s(k)$, therefore it is necessary to integrate over all $k$
at a given frequency $\omega$:
\beq
\gamma_l=\frac{2\omega \mu}{3\rho c_l^3} \sum_k \frac{\omega \tau_s(k)}{1+(\omega \tau_s(k))^2}.
\eeq
Finally, the ratio of the wavelength $\lambda_\omega$ to the mean free path $l_\omega$
is a remarkably simple answer:
\beq
\alpha \equiv \frac{\lambda_\omega}{l_\omega}=
\frac{2}{3} F \frac{c_t^2}{c_l^2},
\label{eqn:ratio} \\
F \equiv \frac{2\pi}{n_0} \int \frac{d^3k}{(2\pi)^3} \frac{\omega \tau_s(k)}{1+(\omega \tau_s(k))^2} \simeq (k_\xi a)^3,
\eeq
where  $n_0=1/a^3$, and
$c_t=\sqrt{\mu/\rho_0}$  is the transverse sound velocity. 
The {\it small}  numerical factor $F$ 
is  {\it a constant, independent of frequency}, since the integrand in the integral is essentially
$\delta$-function  because of the sharp dependence of $\tau_s(k)$ from $k$. 
The constant $F$ is very weakly dependent on the detailed functional form of $\tau_s(k)$
and other parameters.
The simplicity of the main result - Eqn.(\ref{eqn:ratio})-
for the universal ratio gives insight into the origin of universality. This ratio
depends only on the squared ratio of transverse and longitudinal sound velocities and the third power of the ratio of
typical interparticle distance $a$ to the scale of correlated disorder $\xi$. This result is equally
correct for longitudinal and transverse phonons at low frequencies, because the damping is due to the coupling
to short-wavelength
density modes in inhomogeneous media. A more systematic calculation, using kinetic equation
formalism\cite{Landau}, gives the same result. The starting point for such a calculation is that
the low-frequency phonon of frequency $\omega$ causes the variation of the density $\delta \rho_k (\omega)$
at this frequency, for all modes with various $k$'s, and this, in turn, changes the phonon frequency
$\delta \omega= \omega (\gamma_k \delta \rho_k)/\rho_0$, where $\gamma_k$ is a Gruneisen parameter.
Again, it is necessary to account for all modes with various $\tau_s(k)$ disturbed and contributing
relaxationally, since $\omega \tau_s(k) \ll 1$. 


Straightforwardly, the thermal conductivity is $\kappa (T) \simeq C_{v,{\bf ph}}(T) c_0 l(T) \sim T^2$,
where $C_{v,{\bf ph}}(T) \sim T^3$ is the phonon-only specific heat, and for thermal phonons ($\hbar \omega \sim kT$)
the mean free path is $l(T)=\lambda_\omega/\alpha \sim 1/T$.

The density spectrum of Eqn.(\ref{eqn:spectrum}) is essentially given by a {\it linearized} spectrum
of Navier-Stokes equation with visco-elastic moduluses. Various non-linear terms (e.g. of the type 
$v_j \nabla_j v_i$, due to the $\rho v_i v_j$ part of momentum density tensor) should be added
to Eqn.(\ref{eqn:momentum})\cite{Landau} to treat non-linear effects.
 Such non-linearities will modify the harmonic spectrum of slow relaxational modes, 
so that higher energy levels will be inhomogeneously shifted. But the lowest energy levels, given by the linear response,
 can be paired together for any given frequency with distribution of relaxation times (corresponding
to different wavevectors $k$), and, naturally, these lowest energy
levels can be associated with commonly discussed two-level systems (but {\it not local tunneling}
two-level systems!).
Non-linearities are a weak effect in a first approximation if and since the sound attenuation is weak.
 The calculation
of the sound attenuation, which considers coupling between collective modes-sound waves, and
collective relaxational modes-two-level systems, illustrates this point that the interaction between modes is given
by the  small parameter $\alpha$. 
The same parameter $\alpha$ describes both, weak interaction between TLS's and the sound attenuation,
since in the theory presented here TLS's are slow relaxational harmonics of the sound.
This consequence of the theory is consistent
with experiments showing weak interaction between TLS's\cite{Hunklinger}.
Note that echoes were observed on extended collective modes in inhomogeneous media,
examples are plasma and ferrites\cite{Kaplan}. Thus various non-linear effects observed in glasses
(saturation, echoes and spectral diffusion) are not inconsistent with the picture of collective
modes. Non-linear effects as well as long-time logarithmic specific heat
(due to the bulk viscosity) will be addressed in the future.


In this paragraph, I compare briefly the elasto-hydrodynamic theory with theories and proposals suggested
in the literature. The main difference between the Standard Model\cite{Anderson} and the elasto-hydrodynamic theory
is that for the former low-energy excitations are independent tunneling systems, while for the latter
these excitations are collective modes of short-wavelength modulation of the density. But the similarity
between both theories is that the local rearrangements of atoms are assumed to exist.
The shortcomings of the Standard Model
were discussed convincingly, and the long-range dipolar interaction between local defects (e.g. TLS's)
was suggested to lead to universality\cite{Leggett}. In the elasto-hydrodynamic theory, the long-range
interaction between disordered regions (not two-level systems of any sort) is due to elasticity fields. 
The collective nature of TLS's of the present theory is indeed similar in spirit 
 to the earlier proposal\cite{Leggett}. A density spectrum, Eqn.(\ref{eqn:spectrum}), was considered
before\cite{Fulde}, but with a very different assumption about relaxation time $\tau_s$.
It was assumed to be {\it independent of wavevector $k$} and dependent on temperature. Such an assumption
does not lead to linear frequency damping of low-frequency phonons 
(for $\omega \ll \Omega_{BP}$, see Eqn.(\ref{eqn:ratio}))
and does not explain the existence of the Boson peak.

There are several directions how to test experimentally the proposed theory. Direct observation of collective modes
by inelastic neutron and X-ray
scattering at large momentum $k$ and small energies $\omega \ll \Omega_{BP}$ would be a most direct test.
The theory predicts also a weak logarithmic in frequency downward dispersion of sound waves
up to the Boson peak. Another consequence, which follows from Eqn.(\ref{eqn:spectrum}), is the broadening
of sound waves close to the Boson peak, $\Gamma_k \simeq ((c_\infty-c_0)/c_\infty) (c_0 k)^2 \tau_s(k)$,
proportional to $k^2$,
which, it seems, was already observed experimentally\cite{Benassi}, and further stronger broadening due 
to mode-mode coupling. Inelastic X-ray scattering\cite{Benassi} provides a direct support 
for the PA in vitreous silica, for  the strong phonon scattering occurs from
the lengthscale  $2\pi/q_{co} \sim 30 \AA$ (to be associated with $\xi$), which is much longer
than the size of ``elementary structural unit'', $SiO_4$ tetrahedra.
Since the low-frequency collective modes appear as a result of the relaxation of modes around $\Omega_{BP}$,
it would be interesting to test such a connection. 
Experiments with non-linear coupling of spectral diffusion type between the Boson peak modes and two-level systems
(collective modes with large $k$ and small $\omega$) would be desirable. Echo experiments
in their functional relationship to pulse amplitudes and time intervals can differentiate
between local and collective modes\cite{Kaplan}.

It is interesting to inquire further into the origin of universality. Eqn.(\ref{eqn:ratio}) shows
that the small universal number $\alpha$ is given essentially by the cube of the ratio $a/\xi$.
Therefore the universality of $a/\xi$ ratio and the sharp onset of shear relaxation are further
important questions for the investigation.  
The smallness of the ratio $a/\xi$ arises due to the presence of short-range correlation among atoms,
 which can be called medium-range structural order\cite{Elliott}.
Note that the notion of medium-range order does not necessarily mean explicit persistent clusters of atoms.
It seems that the value $\xi/a \sim 6$ (to be consistent with $\alpha \sim 10^{-2}-10^{-3}$)
may be related with the average number of configurational neighbors.
Furthermore, entropic arguments would relate the ratio $a/\xi$ to some fractional power
of the ratio of the glass (or melting) temperature and interatomic bonding energy.
A close analogy of the problem considered here is the problem of a small number associated
with melting transitions. Namely, Lindemann criterion number and small $1/r_s$ ratio
for the melting of electron Wigner crystal ($1/r_s$ is the ratio of kinetic energy to the Coulomb energy
in the many-body electron problem) are small numbers closely related to the ratio $a/\xi$ and describing
in all of these cases short-range coordination (or entropy) in the formation of a crystal or a glass.


In conclusion, from a general assumption of liquid-like properties of glasses
on lengthscales smaller than a certain length $\xi$, the collective low-frequency excitations are derived with 
linear-frequency
density of states $Im\chi_\rho (k,\omega)$ (corresponding to the constant thermodynamic
density of states $P_0 \simeq Im\chi_\rho (k,\omega)/\omega$ of two-level systems, two lowest energy levels of the harmonic
excitations, which are perturbed by non-linear mode-coupling). 
The universality of the sound attenuation ratio $\lambda/l$ is related to the cube of small ratio $a/\xi$
of average interatomic distance $a$ to the medium-range order length $\xi$.

I would like to thank D. Khmelnitskii and A.J. Leggett for valuable  discussions.


\begin{thebibliography}{24}
\bibitem{thermal}
A direct estimate from experimental data shows that the damping of sound
due to thermal conductivity mechanism (in comparison with the viscosity mechanism)
can be neglected at low temperatures in glasses. Essentially because the thermal conductivity
in glasses is significantly smaller than in crystals (where both mechanisms contribute
comparably).
\end{thebibliography}
\end{document}